\documentclass[twocolumn,byrevtex,superscriptaddress,showpacs,preprintnumbers,amsmath,amssymb,aps,pre]{revtex4}
\usepackage{graphicx}% Include figure files
\usepackage{dcolumn}% Align table columns on decimal point
\usepackage{bm}% bold math
\begin{document}
\title{Attempt to distinguish the origins of self-similarity by natural time analysis.}
\author{P. A. Varotsos}
\email{pvaro@otenet.gr}
\affiliation{Solid State Section, Physics Department, University of Athens, Panepistimiopolis, 
Zografos 157 84, Athens, Greece}
\affiliation{Solid Earth Physics Institute, Physics Department, University of Athens, Panepistimiopolis, Zografos 157 84, Athens, Greece}
\author{N. V. Sarlis}
\affiliation{Solid State Section, Physics Department, University of Athens, Panepistimiopolis, Zografos 157 84, Athens, Greece}
\author{E. S. Skordas}
\affiliation{Solid Earth Physics Institute, Physics Department, University of Athens, Panepistimiopolis, Zografos 157 84, Athens, Greece}
\author{H. K. Tanaka}
\affiliation{Earthquake Prediction Research Center, Tokai University 3-20-1, Shimizu-Orido, Shizuoka 424-8610, Japan}
\author{M. S. Lazaridou}
\affiliation{Solid State Section, Physics Department, University of Athens, Panepistimiopolis, Zografos 157 84, Athens, Greece}

\begin{abstract}
Self-similarity may originate from two origins, i.e., the process
memory and the process' increments ``infinite'' variance. A  
distinction is attempted by employing the natural time $\chi$.
Concerning the first origin, we analyze recent data on
Seismic Electric Signals, which support the view that they exhibit
infinitely ranged {\em temporal} correlations. Concerning the second, slowly
driven systems that emit bursts of various energies $E$ obeying
power-law distribution, i.e., $P(E)\sim E^{-\gamma}$ ,are studied.
An interrelation between the exponent
$\gamma$ and the variance $\kappa_1$($\equiv \langle
\chi^2 \rangle - \langle \chi \rangle^2$) is obtained for the shuffled (randomized) data.
In the latter, the most probable value of $\kappa_1$ is approximately equal to
 that of the original data. Finally, it is found that the differential entropy
associated with the probability $P(\kappa_1)$ maximizes for $\gamma$ around $\gamma \approx$ 1.6 to 1.7, which is comparable to the value  determined experimentally in diverse phenomena,
e.g., solar flares, icequakes, dislocation glide in stressed
single crystals of ice e.t.c. It also agrees
with the $b$-value in the Gutenberg-Richter law of
earthquakes.
\end{abstract}
\pacs{05.40.-a, 91.30.Dk, 05.45.Tp, 89.75.-k}
\maketitle

\section{Introduction}
A large variety of natural systems exhibit irregular and complex
behavior which at first looks erratic, but in fact possesses scale
invariant structure (e.g., \cite{PEN95,KAL05}). A process
$\{X(t)\}_{t\geq0}$ is called self-similar\cite{LAM62} if for some
$H>0$,
\begin{equation}
\label{eq1}
X(\alpha t) = \alpha^H X(t)  \quad\quad  \forall\quad  \alpha > 0,
\end{equation}
where the symbol of equality refers here to all finite-dimensional
distributions of the process on the left and the right, and the
parameter $H$ is called self-similarity index or exponent.
Equation (\ref{eq1}) means a ``scale invariance'' of the finite-dimensional
distributions of $X(t)$, which does {\em not} imply, in stochastic
processes, the same for the sample paths (e.g.,\cite{WER05}).
Examples of self-similar processes are Brownian, fractional
Brownian (fBm), L\'{e}vy stable and fractional L\'{e}vy stable
motion (fLsm). L\'{e}vy stable distributions (which are followed
by many natural processes, e.g., \cite{TSA95,TSA96}) differ greatly
from the Gaussian ones because they have heavy tails and their
variance is infinite (e.g.,\cite{WER05,SCA05}).

An important point in analyzing data from natural systems
that exhibit scale invariant structure, is the following: In
several systems this nontrivial structure points to long-range
{\em temporal} correlations; in other words, the self-similarity
results from the process' memory only (e.g., the case of fBm).
Alternatively, the self-similarity may solely result from the
process' increments infinite variance, e.g., L\'{e}vy stable
motion. (Note, that in distributions that are applicable to a large
variety of problems, extreme events have to be {\em truncated} for
physical reasons, e.g., finite size effects, -when there is no
infinity\cite{AUS06}- and this is why we write hereafter ``infinite''.)
 In general, however, the self-similarity may result from
both these origins (e.g., fLsm). It is the main aim of this paper
to discuss how a distinction of the two origins of self-similarity
(i.e., process' memory, process' increments ``infinite'' variance) can
be in principle achieved by employing the natural time analysis.

In a time series comprising $N$ events, the {\em natural time}
$\chi_k = k/N$ serves as an index\cite{NAT01,NAT02} for the
occurrence of the $k$-th event. The evolution of the pair
($\chi_k, Q_k$) is considered\cite{NAT01,NAT02,NAT03,NAT03B,NAT04,NAT05,NAT05B,newbook,VAR05C,VAR06B,TAN04},
where $Q_k$ denotes in general a quantity proportional to the
energy released in the $k$-th event. For example, for dichotomous
signals $Q_k$ stands for the duration of the $k$-th pulse 
while for the  seismicity  $Q_k$ is
proportional to the seismic energy released during the $k$-th
earthquake\cite{NAT01,TAN04,VAR05C} (which is proportional to the seismic moment $M_0$).
 The normalized power spectrum $\Pi(\omega )$ was
introduced\cite{NAT01,NAT02}:
\begin{equation}
\label{eq3}
\Pi (\omega)=\left| \sum_{k=1}^{N} p_k \exp \left( i \omega \frac{k}{N}
\right) \right|^2
\end{equation}
where $p_k=Q_{k}/\sum_{n=1}^{N}Q_{n}$ and $\omega =2 \pi \phi$; 
$\phi$ stands for the {\it natural frequency}. When the system enters into
the {\em critical} stage, the following relation holds\cite{NAT01,NAT02}:
\begin{equation}
\Pi ( \omega ) = \frac{18}{5 \omega^2} -\frac{6 \cos \omega}{5
\omega^2} -\frac{12 \sin \omega}{5 \omega^3}. \label{fasma}
\end{equation}
For $\omega \rightarrow 0$,
Eq.(\ref{fasma}) leads to\cite{NAT01,NAT02,newbook}
$ \Pi (\omega )\approx 1-0.07 \omega^2$
which reflects\cite{VAR05C} that the variance of $\chi$ is given
by $\kappa_1=\langle \chi^2 \rangle -\langle \chi \rangle
^2=0.07$, 
where $\langle f( \chi) \rangle = \sum_{k=1}^N p_k f(\chi_k )$.
 It has been argued\cite{VAR05C}
that in the case of earthquakes, $\Pi(\phi)$  for $\phi \rightarrow 0$, can be considered as
an order parameter and the corresponding probability density
distribution function (PDF) is designated by $P [\Pi(\phi)]$. Since, at $\phi
\rightarrow 0$, $\kappa_1$ is linearly related to $\Pi(\phi)$
(because Eq.(\ref{eq3}) leads to $\Pi(\phi)=1-4 \pi^2\phi^2\kappa_1$ for
$\phi \rightarrow 0$) one can study, instead of $P [\Pi(\phi)]$,
the PDF of $\kappa_1$, i.e.,
$P(\kappa_1)$. This will be used here. 
The entropy $S$ in the natural time-domain is
defined as\cite{NAT01,NAT03B}
$ S \equiv  \langle \chi \ln \chi \rangle - \langle \chi \rangle \ln
\langle \chi \rangle $, which  depends
on the sequential order of events\cite{NAT04,NAT05} and for
infinitely ranged temporal correlations  its
value is smaller\cite{NAT03B,newbook} than the value $S_u (=1/2\ln
2-1/4\approx 0.0966$) of a ``uniform'' distribution (defined in
Refs. \cite{NAT01,NAT03,NAT03B,NAT04,NAT05}, e.g.   when all $p_k$ are
equal), i.e.,$S < S_u$.
The value of the entropy obtained\cite{NAT05B} upon considering
the time reversal ${\cal T}$, i.e., ${\cal T} p_k=p_{N-k+1}$, is
labelled by $S_-$.

This paper is organized as follows: In Section II, we treat the case when 
the self-similarity solely results from the process' memory. Section III deals with  
the self-similarity resulting from the process' increments infinite variance by restricting 
ourselves to slowly driven systems that emit energy bursts obeying power law distributions. 
A brief discussion follows in Section IV, while Section V presents the main conclusion. 

\section{The case of temporal correlations}
\begin{figure}
\includegraphics{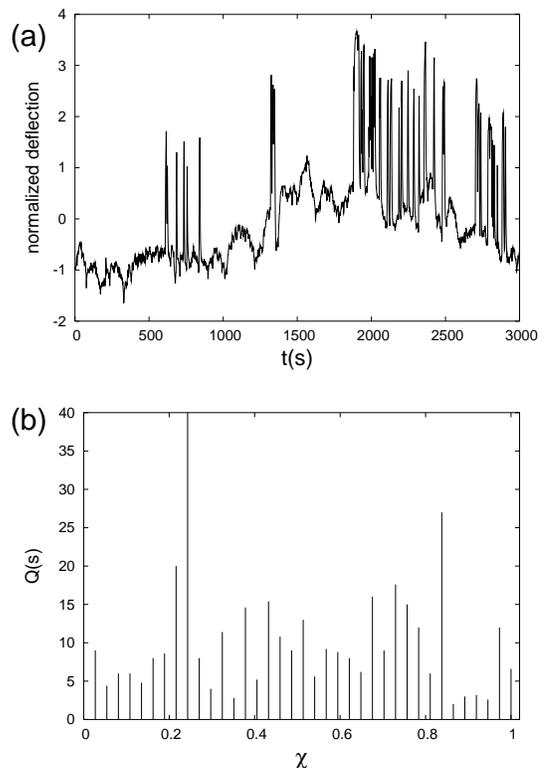}% Here is how to import EPS art
\caption{\label{fg1} (a)An SES activity recently recorded  at PAT station(sampling rate
$f_{exp}=1$Hz). The actual electric field
{\bf E} is 6 mV/km, but here  the signal is presented in normalized
units, i.e., by subtracting the mean value $\mu$ and dividing by
the standard deviation $\sigma$. (b) How the SES activity in (a) 
is read in natural time.}
\end{figure}

We consider, as an example, Seismic Electric
Signals (SES) activities which  exhibit infinitely
ranged temporal correlations\cite{NAT02,NAT03,NAT03B}. 
Figure \ref{fg1}(a) shows a recent
SES activity recorded at a station located in central Greece
(close to Patras city, PAT) on February 13, 2006. It comprises 37
pulses, the durations $Q_k$ of which vary between 1s and 40s 
(see Fig. \ref{fg1}(b)). The natural time representation of this SES activity
can be seen in Fig.\ref{fg1}(b) and the computation of $\kappa_1$,
$S$ and $S_-$ leads to the following values: $\kappa_1 =0.072 \pm 0.002$, 
$S=0.080 \pm 0.002$, $S_- = 0.078 \pm 0.002$. These
values obey the conditions $\kappa_1 \approx 0.070$ and $S, S_-
<S_u$ that have already been found\cite{NAT02, NAT03B, VAR06B} to
be obeyed for other SES activities. If we repeat the computation 
for surrogate data obtained by {\em shuffling} the durations
$Q_{\kappa}$ randomly (and hence their distribution is conserved),
the corresponding quantities, designated by adding a subscript
``shuf'', have the following values: $\kappa_{1,shuf}=0.082$ and
$S_{shuf}(=S_{-,shuf})=0.091$ with standard deviations 0.008 and
0.011 respectively. They are almost equal to the corresponding values of
a ``uniform'' distribution, i.e., $\kappa_u=0.0833$ and
$S_u=0.0966$, although the experimental error in this case is
large due to the small number of pulses. By applying the same 
procedure to other SES activities   reported
earlier\cite{NAT05B}, we find (see Table \ref{tab0}) that actually
$\kappa_{1,shuf} \approx \kappa_u$ and $S_{shuf} \approx
S_{-,shuf} \approx S_u$. This points to the conclusion that the
self-similarity of SES activity results from the process' memory
only, which agrees with an independent analysis of Ref.\cite{WER05}.

\begin{figure}
\includegraphics{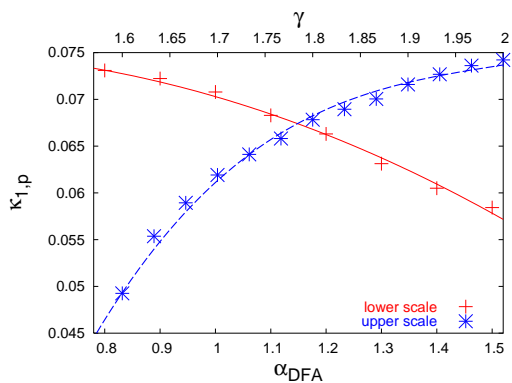}% Here is how to import EPS art
\caption{\label{fg2}(Color online) The values of $\kappa_{1,p}$ as
a function of $\alpha_{DFA}$ for the case of the fBm simulations of 
Ref.\cite{VAR06B}(crosses, lower
scale) or as a function of $\gamma$ for power law distributed data
(asterisks, upper scale). They have been estimated by following
the procedure described in the Appendix B of Ref.\cite{VAR05C}.
The corresponding lines have been drawn as a guide to the eye.}
\end{figure}

\begin{figure}
\includegraphics{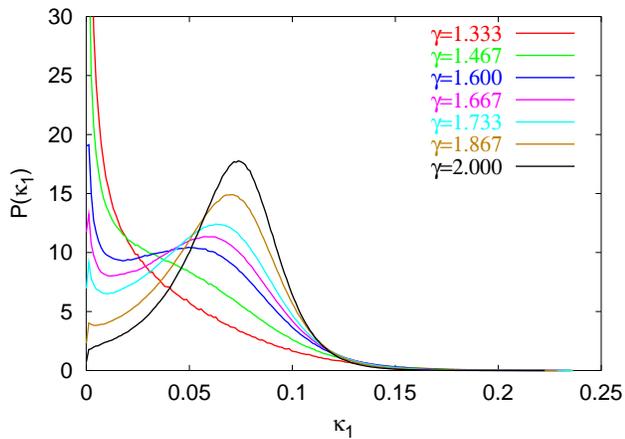}% Here is how to import EPS art
\caption{\label{fg3} (Color online) The probability
density function $P(\kappa_1)$ versus $\kappa_1$ for several
values of $\gamma$ (see the text and \cite{VAR}).}
\end{figure}

\begin{figure}
\includegraphics{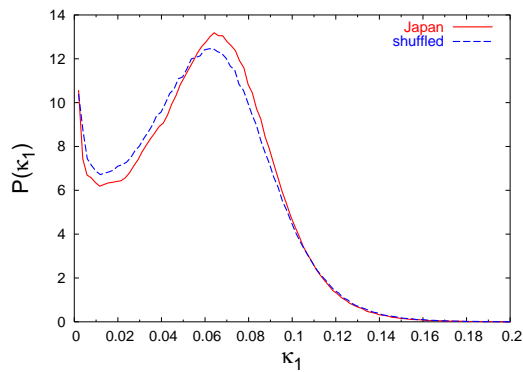}% Here is how to import EPS art
\caption{\label{fg4} (Color online)The PDFs of $\kappa_1$ when
using either the actual seismic catalogue of Japan(solid) treated in Ref.\cite{VAR05C} or  the same data in random order (dashed).}
\end{figure}

\begin{figure}
\includegraphics{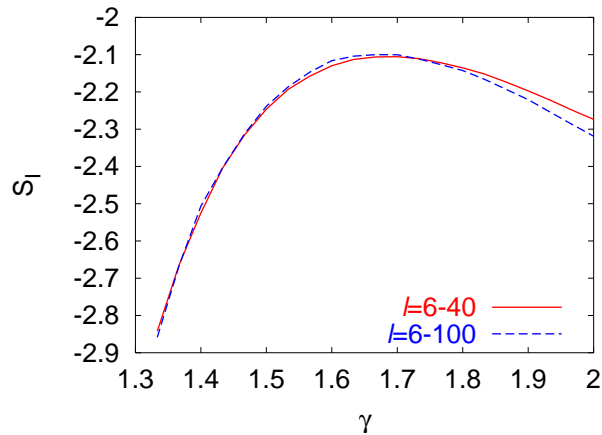}% Here is how to import EPS art
\caption{\label{fg5} (Color online) The calculated values of the
differential entropy $S_I$ versus the exponent $\gamma$. Two window lenghts are used and their 
results are almost the same.}
\end{figure}

\begin{table*}
\caption{ The values of $\kappa_1$, $\kappa_{1,shuf}$, $S$, $S_-$
and  $S_{shuf}$ for the SES activities
 mentioned in Ref.\cite{NAT05B} as well as the one (PAT) depicted in Fig.\ref{fg1}.
  The numbers in parentheses
  denote the standard deviation for the distributions of  $\kappa_{1,shuf}$ and
 $S_{shuf}$ in the shuffled data. } \label{tab0}
\begin{ruledtabular}
\begin{tabular}{cccccc}

  Signal & $\kappa_1$ & $\kappa_{1,shuf}$               & $S$ &
   $S_-$ & $S_{shuf}$\footnotemark[2] \\
\hline
K1  & 0.063$\pm$0.003\footnotemark[1] & 0.083(0.005) & 0.067$\pm$0.003\footnotemark[1] & 0.074$\pm$0.003\footnotemark[1] & 0.096(0.007) \\
K2  & 0.078$\pm$0.004\footnotemark[1] & 0.082(0.007) & 0.081$\pm$0.003\footnotemark[1] & 0.103$\pm$0.003\footnotemark[1] & 0.094(0.009) \\
A   & 0.068$\pm$0.004\footnotemark[1] & 0.082(0.007) & 0.070$\pm$0.008\footnotemark[1] & 0.084$\pm$0.008\footnotemark[1] & 0.092(0.010) \\
U   & 0.071$\pm$0.004\footnotemark[1] & 0.082(0.009) & 0.092$\pm$0.004\footnotemark[1] & 0.071$\pm$0.004\footnotemark[1] & 0.093(0.012) \\
T1  & 0.084$\pm$0.007\footnotemark[1] & 0.082(0.007) & 0.088$\pm$0.007\footnotemark[1] & 0.098$\pm$0.010\footnotemark[1] & 0.091(0.010) \\
C1  & 0.074$\pm$0.002\footnotemark[1] & 0.082(0.008) & 0.083$\pm$0.004\footnotemark[1] & 0.080$\pm$0.004\footnotemark[1] & 0.092(0.011) \\
P1  & 0.075$\pm$0.004\footnotemark[1] & 0.082(0.008) & 0.087$\pm$0.004\footnotemark[1] & 0.081$\pm$0.004\footnotemark[1] & 0.090(0.011) \\
P2  & 0.071$\pm$0.005\footnotemark[1] & 0.082(0.009) & 0.088$\pm$0.003\footnotemark[1] & 0.072$\pm$0.015\footnotemark[1] & 0.091(0.012) \\
E1  & 0.077$\pm$0.017\footnotemark[1] & 0.083(0.008) & 0.087$\pm$0.007\footnotemark[1] & 0.081$\pm$0.007\footnotemark[1] & 0.092(0.010) \\
PAT & 0.072$\pm$0.002 & 0.082(0.008) & 0.080$\pm$0.002 &
0.078$\pm$0.002 & 0.091(0.011)
\end{tabular}
\footnotetext[1]{from Ref.\cite{NAT05B}} \footnotetext[2]{note
that $S_{shuf}=S_{-,shuf}$ as mentioned in the text.}
\end{ruledtabular}
\end{table*}

%\begin{turnpage}
%\squeezetable
\begin{table*}
\caption{Compilation of the experimental values of the power-law
exponent $\gamma$ determined in different processes.} \label{tab1}
\begin{ruledtabular}
\begin{tabular}{ccc}
Process / type of measurement & $\gamma$ & References \\
\hline
Dislocation glide in hexagonal  & 1.6 & \cite{MIG01}\\
ice single crystals (Acoustic emission) & & \\
& & \\
Solar flares & 1.5 - 2.1 & \cite{ASC00,PAR00,HUG03,NIG04}\\
& & \\
Microfractures before the  & 1.51 & \cite{GAR97, GUA02}\\
breakup of wood (acoustic emission) & & \\
& & \\
Microfractures before the  & 2.0 & \cite{GAR97, GUA02}\\
breakup of fiberglass (acoustic emission) & & \\
& & \\
Earthquakes & 1.5 - 1.8 ($b$ = 0.8 to 1.2) & See Ref.\cite{RUN03} \\
& & and references therein\\
& & \\
Icequakes & $\sim$1.8 ($b \approx 1.25$) & See p.212 of \cite{WEI03} \\
& &  and references therein

\end{tabular}
\end{ruledtabular}
\end{table*}

In addition, the Detrended Fluctuation Analysis
(DFA)\cite{PEN94,BUL95}  in natural time of the SES activity depicted in
Fig.\ref{fg1}, leads to an exponent
$\alpha_{DFA}=1.07 \pm 0.36$, which agrees with the earlier
finding $\alpha_{DFA} \approx 1$ in several other SES
activities\cite{NAT03}. Interestingly, the values of the
quantities $\kappa_1, S, S_-$ and $\alpha_{DFA}$ are consistent
with the results deduced from a numerical simulation in fBm
time series described in Ref.\cite{VAR06B}; the latter showed
that when $\alpha_{DFA} \approx 1$ the corresponding values are:
$\kappa_1 \approx 0.070$, $S \approx S_- \approx 0.080$. Figure \ref{fg2} 
depicts the most probable value $\kappa_{1,p}$ of $\kappa_1$
 versus
$\alpha_{DFA}$ resulting from such  a numerical simulation.

\section{the case of power law distributions}
We now study a case of self-similarity
resulting from the process' increments ``infinite'' variance.
 Here, we restrict ourselves to slowly driven systems that emit
energy  bursts obeying power law distribution
\begin{equation}
P(E)\sim E^{-\gamma}
\label{eq9}
\end{equation}
where $\gamma$ is constant. In a large variety of such systems, in
diverse fields, an inspection of the recent experimental data
reveals that the $\gamma$ exponent lies in a narrow range, i.e.,
$1.5 \leq \gamma \leq 2.1$ (and mostly even within narrower
bounds, i.e., $\gamma =$ 1.5 to 1.8). To realize the diversity of
the phenomena that exhibit the aforementioned property, we compile
some indicative examples in Table \ref{tab1}, which are the
following. 

First, crystalline materials subjected to an external
stress, display bursts of activity owing the nucleation and motion
of dislocations. These sudden local changes produce acoustic
emission waves which reveal that a large number of dislocations
move cooperatively in an intermittent fashion (e.g., see
\cite{KOS04} and references therein). As a precise example, we
include in Table \ref{tab1} the results of acoustic emission
experiments on stressed single crystals of ice under viscoelastic
deformation (creep), which show that the probability distribution
of energy bursts intensities obey a power-law distribution with
$\gamma =$ 1.6 spanning many decades (see Fig.\ref{fg1} of
\cite{MIG01}). Second, we consider the case of solar flares that
represent impulsive energy releases in the solar corona (e.g. see
Ref. \cite{NIG04} and references therein; see also
Ref.\cite{ARC06} in which it is concluded that earthquakes and
solar flares exhibit the same distributions of sizes,
interoccurrence times, and the same temporal clustering). This
energy release is observed in various forms: thermal, soft and
hard x-ray emissions, accelerated particles etc. The statistical
analysis of these impulsive events show that the energy
distribution exhibit, over several orders of magnitude, a
power-law with exponents $\gamma$ ranging from 1.5 to
approximately 2.1 (depending on the experimental procedure and the
geometrical assumptions adopted in the analysis). Other examples
are: acoustic emission from microfractures before the breakup of
heterogeneous materials (wood, fiberglass), icequakes and
earthquakes.
%To realize the diversity of
%the phenomena that exhibit the aforementioned property, we compile
%some indicative examples in Table \ref{tab1}.  These are the
%following: Acoustic emission experiments on stressed single crystals of ice 
%that  show energy bursts (see Fig.1 of
%\cite{MIG01}) arising from a large number of dislocations
%that move cooperatively in an intermittent fashion\cite{KOS04},  solar flares that
%represent impulsive energy releases in the solar corona\cite{NIG04,ARC06}, 
%acoustic emission from microfractures before the breakup of
%heterogeneous materials (wood, fiberglass), icequakes and
%earthquakes. 

Concerning the latter,  the
best known scaling relation is the
Gutenberg-Richter law\cite{GUT54}, which states that the
(cumulative) number of earthquakes with magnitude greater than m
occurring in a specified area and time is given by
\begin{equation}
N(>m)\sim 10^{-bm}
\label{eq10}
\end{equation}
where $b$ is a constant, which varies only slightly from
region to region (cf. Eq.(\ref{eq10}) holds both {\em regionally} and {\em
globally}) being generally in the range $0.8\leq b \leq 1.2$ (see
\cite{RUN03} and references therein). Considering that the seismic
energy $E$ released during an earthquake is related\cite{KAN78} to the
magnitude through $E \sim 10^{cm}$ -where $c$ is around
1.5- Eq.(\ref{eq10}) turns to Eq.(\ref{eq9}), where $\gamma = 1 + b/1.5$. Hence, 
$b \approx 1$ means that the exponent $\gamma$ is around $\gamma$=1.6
to 1.7.

The following procedure is now applied: We generate(see
also\cite{VAR}) a large amount
of artificial data obeying Eq.(\ref{eq10}) for a certain $\gamma$ value. 
These {\em randomized}
(``shuffled''\cite{NAT04}) data are subsequently analyzed, in the
natural time domain, for each $\gamma$ value, with the following
procedure\cite{VAR05C}: First, calculation of
the variance $\kappa_1$ is made for an event taking time windows
for 6 to 40 consecutive events (the choice of the precise value of
the upper limit is not found decisive, because practically the
same results are obtained even if the number of consecutive events
was changed from 6 - 40 to 6 - 100, as it will be further
discussed below). And second, this process was performed for all
the events by scanning the whole dataset. In Fig.\ref{fg3} we plot
the quantity $P(\kappa_1)$ versus $\kappa_1$ for several
$\gamma$-values. The most probable value $\kappa_{1,p}$ (for
$\gamma$ =constant) is also plotted in
Fig.\ref{fg2} versus the corresponding $\gamma$-value. This curve
interrelates $\kappa_1$ and $\gamma$ for the shuffled data (thus an
eventual process' memory is here destroyed\cite{NAT04}) and hence the plotted
$\kappa_{1,p}$ values (which differ markedly from $\kappa_u$)
correspond to the self-similarity resulting from the heavy-tailed
distribution only.

In order to identify the origin of
self-similarity in a real data set, let us consider here the example
of earthquakes. Using the Japan catalogue mentioned in
Ref.\cite{VAR05C}, we give in Fig.4 the two curves $P(\kappa_1)$
versus $\kappa_1$ that result when the aforementioned calculation is made
by means of a window of 6-40 concecutive events sliding through either the
original catalogue or a shuffled one. Comparing the resulting
$\kappa_{1,p}$ values (both of which markedly differ from
$\kappa_u$) we see that the value of the surrogate data ($\approx 0.064$)
differs slightly from the one ($\approx 0.066$) corresponding to the
original data. This reflects that the self-similarity mainly
originates from the process' increments ``infinite'' variance.
Note, however, that the $\kappa_{1,p}$ value of the original data is
comparable to the value $\kappa_1$=0.070 that was found in 
infinitely ranged {\em temporal} correlations. This merits further
investigation.

\section{Discussion}
Here, we  discuss a challenging  point that emerges from a further
elaboration of the results depicted in Fig.3. First, note that upon
increasing the $\gamma$ value from $\gamma$ = 1.3 to 2.0, the
feature of the curve changes significantly, becoming bimodal at
intermediate $\gamma$-values. Second, we calculate, for each
$\gamma$-value studied, the so called {\em differential entropy},
defined as $S_I=-\int P(\kappa_1)\ln P(\kappa_1)d\kappa_1$ which is
the Shannon information entropy of a continuous probability
distribution, e.g., see \cite{GAR04}. (Note that, the Shannon
information entropy is {\em static} entropy and {\em not} a
dynamic one \cite{NAT04}.) Finally, we investigate the resulting
$S_I$-values versus $\gamma$. Such a plot is given in Fig.5, whose
inspection reveals that $S_I$ maximizes at a value of $\gamma$
lying between $\gamma$ = 1.6 and $\gamma$=1.7, which is more or
less comparable with the experimental values, see Table
\ref{tab1}. (This value is not practically affected by the 
window length ($l$) chosen; in reality, upon increasing $l$
from $l$=10 to $l$=1000, we find that, $\gamma$-value at which
$S_I$ maximizes in Fig.\ref{fg2}, decreases only slightly from
$\gamma$=1.70 to $\gamma$=1.63.) In particular for the case of
earthquakes this $\gamma$-value corresponds to $b \approx 1$, thus
agreeing with the experimental findings mentioned above. Does it
mean that the $b$ or $\gamma$ value can be determined just by
applying the Maximum Entropy Principle in the sense developed by
Jaynes\cite{JAY57,JAY03}, who suggested to look statistical mechanics as
a form of {\em statistical inference} and {\em start} statistical
physics from the principle of maximum entropy inference (MaxEnt)?
This is not yet clear, because a widely
accepted formalism for {\em non-equilibrium} statistical
mechanics is still lacking. 

Finally,  the fact that in some experiments the
resulting $\gamma$-values differ slightly from $\gamma$=1.6 to 1.7
predicted from Fig.\ref{fg5}  could be attributed to the following: 
Figure \ref{fg5} is based on randomized data, while the actual
data may also exhibit temporal correlations (e.g., the case
of aftershocks). In addition, finite size effects\cite{AUS06} might play 
a significant role.

\section{Conclusion}
In summary, the origin of
self-similarity may be distinguished as follows: If self-similarity
exclusively results from the process' memory, the $\kappa_1$ value
should {\em change} to $\kappa_u$=0.0833 (and the values of $S, S_-$ to
$S_u$=0.0966) for the surrogate data. On the other hand, if the
self-similarity results from process' increments ``infinite''
variance only, the $\kappa_{1,p}$ values should be the same (but differing from $\kappa_u$) for the
original and surrogate data.

When studying the differential entropy associated with the PDF of $\kappa_1$  it maximizes when
the exponent $\gamma$ lies in the narrow range 1.6 to 1.7, in agreement with the experimental findings in
diverse fields. This, for the case of earthquakes, immediately
reflects that the $b$-value in the Gutenberg-Richter law is
$b \approx 1$, as actually observed.

%\end{squeezetable}
%\end{turnpage}

%\bibliography{bval}
%\bibliographystyle{apsrev}

\end{document}